\documentclass{article}

\begin{document}

\def\giorno{10/11/2001}

\def\Fb{{\bf F}}
\def\F{{\cal F}}
\def\L{{\cal L}} 
\def\M{{\cal M}} 
\def\a{\alpha}
\def\b{\beta}
\def\eps{\varepsilon}
\def\de{\delta}
\def\om{\omega}
\def\Om{\Omega}
\def\La{\Lambda}
\def\sse{\subseteq}
\def\ss{\subset} 
\def\s{\sigma}
\def\S{\Sigma} 
\def\ga{\gamma}
\def\toro{{\bf T}}
\def\T{{\rm T}}
\def\de{\delta}
\def\d{{\rm d}}
\def\phi{\varphi}
\def\pa{\partial}
\def\w{\wedge}
\def\({\left(}
\def\){\right)}
\def\[{\left[}
\def\]{\right]}
\def\^#1{{\widehat #1}}
\def\~#1{{\widetilde #1}}

\def\sp{\medskip \par\noindent}
\def\EOP{$\triangle$ \par \medskip}
\def\EOR{$\odot$ \par \medskip}

\def\interno{\hskip 2pt \vbox{\hbox{\vbox to .18
truecm{\vfill\hbox to .25 truecm
{\hfill\hfill}\vfill}\vrule}\hrule}\hskip 2 pt}

\def\mapright#1{\smash{\mathop{\longrightarrow}\limits^{#1}}}
\def\mapdown#1{\Big\downarrow\rlap{$\vcenter{\hbox{$\scriptstyle#1$}}$}}
\def\mapleft#1{\smash{\mathop{\longleftarrow}\limits^{#1}}}
\def\mapup#1{\Big\uparrow\rlap{$\vcenter{\hbox{$\scriptstyle#1$}}$}}


\title{The Poincar\'e--Lyapounov--Nekhoroshev theorem }

\author{Giuseppe Gaeta\footnote{Supported by ``Fondazione CARIPLO per la Ricerca Scientifica'' under project ``Teoria delle perturbazioni per sistemi con simmetria''. E-mail: gaeta@berlioz.mat.unimi.it or gaeta@roma1.infn.it} \\ 
{\it Dipartimento di Matematica, Universit\'a di Milano} \\
{\it via Saldini 50, I--20133 Milano (Italy)} }

\date{\giorno}

\maketitle

\noindent
{\bf Summary.} {We give a detailed and mainly geometric  proof of a theorem by N.N. Nekhoroshev for hamiltonian systems in $n$ degrees of freedom with $k$ constants of motion in involution, where $1 \le k \le n$. This states persistence of $k$-dimensional invariant tori, and local existence of partial action-angle coordinates, under suitable nondegeneracy conditions. Thus it admits as special cases the Poincar\'e-Lyapounov theorem (corresponding to $k=1$) and the Liouville-Arnold one (corresponding to $k = n$), and interpolates between them. The crucial tool for the proof is a generalization of the Poincar\'e map, also introduced by Nekhoroshev.}

\bigskip\bigskip\bigskip\bigskip

\section*{Introduction.}

In the early nineties, Nekhoroshev obtained an interesting result on
persistence of tori in hamiltonian systems having ``partial integrability'', i.e. a number of integrals of motion greater than one but smaller than the number of degrees of freedom \cite{Nek1}. 

As pointed out by Nekhoroshev, this result represents a bridge between the Poincar\'e-Lyapounov theorem (persistence of periodic trajectories) and the Liouville-Arnold one (complete integrability). 
Indeed, on the one hand it extends the Poincar\'e-Lyapounov theorem (which applies to invariant curves which are topologically $S^1$ circles) to the case of higher dimensional tori $\toro^k = S^1 \times ... \times S^1$; and on the other it extends the Liouville-Arnold theorem by showing that one can build ``partial action-angle coordinates'' on these invariant tori, i.e. one can define coordinates $(I_\a, \phi^\a ; p_i,q^i)$, with $\a = 1,...,k$, $i=1,...,n-k$, in a open set $N \ss M$ (see below) so that the invariant tori correspond to $p=q=0$ and are parametrized by the value of the action coordinates $I_\a$. 
 
Unfortunately a detailed proof of Nekhoroshev's theorem has never been
published. Here we want to provide such a detailed proof, which presumably is not too different from the one considered by Nekhoroshev: we use indeed a construction
introduced by him to state the needed nondegeneracy condition, which generalizes the Poincar\'e one (based on the spectrum of the linearization of the Poincar\'e map); we call the map on which this nondegeneracy condition is based, the {\it Poincar\'e-Nekhoroshev map}.  

This construction is considered in section 1, while section 2 is devoted to the actual proof of the result by Nekhoroshev; more precisely, we limit our discussion to the existence of a fibration in tori: once this has been obtained, the construction of action-angle coordinates is quite standard and can be obtained e.g. by just following the Arnold discussion \cite{Arn1} for the case of complete integrability in tori ($k=n$), so that we omit it.

It should be stressed that one could also -- as we briefly discuss below -- partially extend Nekhoroshev's result to the non-hamiltonian setting; needless to say, such extension will be limited to the assertions on fibration in tori of a given region of the phase space and will not encompass the part of Nekhoroshev's theorem relating to action-angle variables. 

In this note we pursue a geometric (as far as possible) approach, following the original discussion by Nekhoroshev. However, once the geometry of the problem has been understood, one can accordingly define convenient local coordinates in the neighbourhood of any point on the invariant torus $\Lambda$; in these the study of the Poincar\'e-Nekhoroshev map -- and thus of all the questions dealt with 
here -- results easier from the analytical point of view. Such an approach is followed in \cite{BG}, where it is also discussed how the algebraic approach to the nondegeneracy condition discussed in section 6 below applies in a perturbative frame to the study of the dynamics around quite general invariant tori; the same approach can also be applied to study quasiperiodic breathers in certain classes of infinite dimensional systems \cite{BV}.
 
\bigskip

The {\bf plan of the paper} is as follows. In section 1 we state precisely the result to be proved.  In section 2 we fix some notation, describe some geometrical construction (a double local foliation, which is later proved to be a local fibration), and by means of this we define the Poincar\'e-Nekhoroshev map.  In section 3 we state the main nondegeneracy condition (condition N) and apply the tools developed in section 1 to discuss the existence of invariant tori nearby the known one. The discussion  of section 3 depends on some arbitrary choice, and in section 4 we study if and how the results obtained do actually depend on these choices. The later sections are devoted to a detailed study of several examples, focusing -- for obvious interest in applications -- on perturbed oscillators: in section 5 we deal with symmetrically perturbed oscillators, i.e. on systems for which the integrals of motion are explicitely known, and in section 6 we consider the case where the perturbation preserves some integrals of motions which are not explicitely known; in these cases the nondegeneracy condition has a very convenient algebraic  formulation. In the Appendix we briefly discuss the extension of Nekhoroshev's result to the non-hamiltonian case.

\subsection*{Acknowledgements}

I would like to thank N.N. Nekhoroshev for calling my attention to his theorem discussed here. This note grew out of discussions with F. Cardin and especially with D. Bambusi; I would like to thank both of them.
The financial support of {\it ``Fondazione CARIPLO per la Ricerca Scientifica''} under project {\it ``Teoria delle perturbazioni per sistemi con simmetria''} is gratefully acknowledged. 

\section{The Poincar\'e-Lyapounov-Nekhoroshev \\ theorem.}

We will first of all state precisely the result we want to prove in detail \cite{Nek1}. In order to simplify the expression of the nondegeneracy condition $(iii)$ we will refer to the {\it Poincar\'e-Nekhoroshev map}, to be introduced and studied in section 2 below.

\sp
{\bf Theorem (Nekhoroshev).} {\it Let $(M,\om )$ be a $2n$-dimensional symplectic manifold. Let $\Fb := \{ F_1 , ... , F_k \}$ be $k$ real
functions on $M$, differentiable of class $C^r$, such that their associated hamiltonian flows $X_i$ satisfy $[X_i , X_j ] = 0 $ globally on $M$, for all $i,j \in \{ 1,...,k \}$. 

Assume moreover that: {\rm (i)} there exists a $C^r$ $k$-dimensional torus $\La \ss M$ invariant under all the $X_i$, and that $\d F_i \wedge ... \wedge \d F_k \not=0 $ on all points of $\La$; {\rm (ii)} there is a $c \in R^k$ such that the vector field $X_c = \sum_i c_i X_i$ has closed trajectories with period 1 on $\La$; {\rm (iii)} the spectrum of the linear part $A_c$ of the Poincar\'e-Nekhoroshev map associated to $X_c$ does not include the unity. 

Then, in a neighbourhood $U$ of $\La$ in $M$, there is a symplectic submanifold $N$ which is fibered over a domain $B \ss R^k$ with as fibers $C^r$ differentiable tori $\La_\b = N \cap \Fb^{-1} (\b) \simeq \toro^k$ (with $\b \in B$) $C^r$-diffeomorphic to $\La$ and invariant under the $X_i$.}
\bigskip

This theorem represents a generalization of the Poincar\'e-Lyapounov theorem on invariant $S^1$ orbits to invariant $\toro^k$ manifolds (obviously $S^1$ and $\toro^k$ should be meant in topological or $C^r$ sense), and thus should be called the {\it Poincar\'e-Lyapounov-Nekhoroshev theorem} (PLN theorem). 

Actually, Nekhoroshev paper \cite{Nek1} included two other statements, concerning the construction of action-angle coordinates in $N$; these were a generalization of the Liouville-Arnold theorem on the construction of action-angle variable for $\toro^n$ to lower dimensional tori $\toro^k$. This second part could be called the Poincar\'e-Arnold-Nekhoroshev theorem. We do not discuss the proof of these statements on action-angle variables, as their proof is just the same as in the completely integrable case $k=n$, thanks to lemma 4; see e.g. the discussion in Arnold's book \cite{Arn1}. 

Similarly, in Nekhoroshev's original statement of the theorem $\La$ was just assumed to be compact and connected, but the proof that in this case $\La$ is just a torus is standard; see again \cite{Arn1}.

\section{The Poincar\'e-Nekhoroshev map.}

\subsection{General setting}

Let $(M,\om )$ be a symplectic manifold (with symplectic form $\om$) of dimension ${\rm dim} (M) = 2n$ and differentiable of class $C^r$ ($r \ge 1$). Let $\Fb := \{ F_1 , ... , F_k \}$ be $k$ independent real functions on $M$, also differentiable of class $C^r$, such that their associated hamiltonian flows $X_i$ (defined by $X_i \interno \om = \d F_i$) are linearly independent on $\La$ and satisfy $[X_i , X_j ] = 0 $ for all $i,j \in \{  1,...,k \}$.

In this case we can consider flows under any hamiltonian vector field $X_c = \sum c_i X_i$ (with $c_i \in R$), and all of these will have $k$ integrals of motion in involution, i.e. the $F_i$'s. 

It is well known that if $k=n$ this leads to affirm integrability of the hamiltonian system. Here we consider the case where $k < n$; note that for $k=1$ we will be in the setting of the Poincar\'e-Lyapounov theorem.

We will {\it assume} that there is a compact and connected manifold $\Lambda \ss M$ of dimension $k$, differentiable of class $C^r$, which is invariant under all the $X_i$, and such that the $X_i$ are linearly independent at all points $m \in \Lambda$. Obviously this $\Lambda$ will be a submanifold of $\Fb^{-1} (\beta_0)$ for some $\beta_0 \in R^k$; we will denote this common level manifold for the $F_i$ as $\F$.
 
It is actually immediate to observe that, being $k$-dimensional, connected, compact and invariant under the $k$ commuting vector fields $X_i$, the $C^r$ manifold $\La$ is necessarily a torus: $\La = \toro^k$ (see e.g. the proof of the Liouville-Arnold theorem in \cite{Arn1}). 

Thus, under our assumption the system has an invariant $k$-torus. It is well known that for $k=1$ (so that $\La \approx S^1$ is a periodic orbit) and under a nondegeneracy condition which can be stated in terms of the Poincar\'e map, this is part of a one-parameter local family of such orbits (Poincar\'e-Lyapounov theorem). 
We want to show that the same holds for arbitrary $k$, $1 \le k \le n$; the main problem lies in identifying the appropriate nondegeneracy
condition.
 
\sp
{\bf Remark 1.} In order to study invariant tori by analytical and/or geometrical (not numerical) means, it is not convenient to use the Poincar\'e map: first of all, if the flow on the torus is irrational the Poincar\'e map would not have a fixed point, and our argument below based on a fixed point theorem would not work; moreover, if we have a periodic flow, around any point of the invariant torus we have a trivially invariant $(k-1)$-manifold for the Poincar\'e map. \EOR
 
We would thus like to have some equivalent of the Poincar\'e map which 
removes this degeneracy, i.e. some kind of ``Poincar\'e map modulo torus action''.
This is provided by the {\it Poincar\'e-Nekhoroshev map}, introduced by Nekhoroshev \cite{Nek1}, which we illustrate here.

\subsection{The first foliation}

For ease of language, we take angular coordinates $(\phi_1 , ... , \phi_k )$ on $\La$; we denote points on $\La$ as $m$ or $m(\phi)$ when we want to use these coordinates. 
Our considerations will be based at a given but arbitrary point $m_0 = m (\phi_0) \in \La$. 

We take a (local, for $\phi$ in a neighbourhood ${\cal Z}_0$ of $\phi_0$) $k$-parameter family of local $(2n-k)$-manifolds $\S_\phi$, differentiable of class $C^r$, passing through $m (\phi)$ and transversal to both $\La$ and $\F$ in $m(\phi)$; we also write $\S_m$ rather than $\S_\phi$, to emphasize this is the manifold passing through the point $m (\phi) \in \La$, and denote by ${\widehat U} \ss M$ the neighbourhood of $m_0 \in \La$ where this family of local manifolds is defined; obviously ${\cal Z}_0 = {\widehat U} \cap \La$. 

As $\Fb$ is invariant under all the $X_i$, it can be helpful -- specially in trying to visualize the construction -- to consider the restriction of our setting to $\F$. We denote by $\s_\phi$ the intersection $\s_\phi = \S_\phi \cap \F$; obviously the $\s_\phi$ are transversal to $\La$ in $\F$.

\sp
{\bf Remark 2.} We note that the linear independence of the $X_i$ at all points of $\La$ can be restated, as the $X_i$ are obtained from the $F_i$ through the action of the symplectic form, by saying that $ \d F_1 \wedge ... \wedge \d F_k \not= 0$ at all points of $\La$ (which also guarantees independence of the $F_i$ in a neighbourhood of  $\La$); this guarantees we can actually have local manifolds transversal to the foliation of $M$ given by common level sets of $\Fb$, and that the $\S_m$ will actually be such. \EOR

The distribution $\S_\phi$ defines a partition of the neighbourhood ${\widehat U} \ss M$ of $m_0 \in \La$ into equivalence classes. 
We also write ${\widehat V} = {\widehat U} \cap \F$; the distribution
$\s_\phi$ defines a partition of ${\widehat V}$ into equivalence classes. 
We anticipate that these partitions, which define a local foliation, will be the basis of our construction: it will result that they define actually a global foliation, and we want then to build a fibration out of it. 

\sp
{\bf Remark 3.} Were we considering a manifold equipped with a metric, this construction would be simpler, as one could consider the normal bundle of $\La$ as a building ground for it; see also the appendix. \EOR

\subsection{Geometry around the invariant torus: \\ the second foliation.}

Together with the foliation of $U$ given by the $\S_m$, we consider a local foliation $\L^{(m_0)}$ (which we also denote simply by $\L$ for ease of notation), defined in a neighbourhood $\~W$ of $m_0$, associated to the fields $X_i$. 

We denote by $L^{(m_0)}_p$ (or simply as $L_p$) the leaf of $\L^{(m_0)}$ passing through the point $p \in \~W$. The leaves of $\L^{(m)}$ are built as follows: at any point $x \in \~W$ the vector  fields $X_i$ define a hyperplane $\chi_x \approx R^h$ ($h \le k$) in $\T_x M$; this distribution defines local manifolds tangent to the $\chi_x$ in the neighbourhood of any point $y \in M$ and -- for $y \in \~W$ -- these are the leaves of $\L^{(m_0)}$. 

Note that necessarily each leaf of this foliation is a submanifold of some $\Fb^{-1}$, so that it restricts naturally to a local foliation $\L^{(m_0)}_\F $ of the neighbourhood $\~w = \~W \cap \F$ in $\F$, with
leaves $L^{(m_0)}_\F = L^{(m_0)} \cap \F$.  

The condition of linear independence of the $X_i$ on $\La$ guarantees that in
a neighbourhood ${\widetilde U} \sse \~W$ of $\La$ (we write ${\widetilde V} = {\widetilde U} \cap \F$) the hyperplanes $\chi_x$ are of constant dimension
$k$. The same applies therefore to the leaves of $\L^{(m)}$, and to the
intersection of these with $\F$.
 
Finally, we denote $W := {\widehat U} \cap {\widetilde U}$, $w = W \cap \F$.
From now on we will, for ease of language, take $\S_m$ and $\L^{(m_0)}$ to be
defined only in $W \equiv W_{m_0}$. 
 
In $W$ (respectively, in $w$), both the foliations are well defined, with leaves $\S_m$ and $\L^{(m_0)}_p$ (respectively, $\s_m$ and $[L^{(m_0)}_\F]_q$) of constant dimensions: $\dim (\S_m) = 2n - k$, $\dim (\L^{(m_0)}_p) = k$ (respectively, $\dim( \s_m) = 2n - 2k$ and $\dim( [L^{(m_0)}_0]_q) = k$). 

Thus, for any point $p \in W \cap \S_m$ (resp. any point $q \in w \cap \s_m$), we have $\S_m \cap \L_p = \{ p \}$ (resp., $\s_m \cap L_q = \{ q \}$). That is, these are transverse local foliations in $W$ (resp. in $w$), and the intersections of leaves reduce to a single point.

\sp
{\bf Remark 4.} In this way we have defined a local (trivial) fibration ${\cal E}_0 = (\pi_0 : W_{m_0} \to {\cal Z}_0$ in $W \equiv W_{m_0}$, with fibers $\pi^{-1} (m) = \S_m$. Note that the fields $X_i$ can be taken as a distribution of horizontal vector fields in ${\cal E}_0$, i.e. they define a connection. \EOR
 
\sp
{\bf Remark 5.} In the neighbourhood $W \equiv W_{m_0}$ of the point $m_0 \in \La$, we can and will take a local system of coordinates $(\phi,s,F)$, where $\phi$ and $F$ are as above, and $s= (s_1 , ... , s_{2n-2k} )$ can be chosen as local coordinates along the $\s_m$ [obviously $m_0$ has coordinates $(\phi_0 , F (\beta_0), 0)$]. Indeed, the foliation $\L^{(m_0)}$ provides a way to compare coordinates pertaining to different $\s_m$ manifolds: as we have shown, each leaf meets each $\s_m$ in one and only one point. Needless to say, this remark and the previous one are the same observation in different languages; this coordinates formulation will be used in lemma 1 below. \EOR
 
Note that we can repeat this construction with an arbitrary given point $m \in \La$ as base point; we denote the corresponding neighbourhoods and foliations as  $W_m \ss M$ (resp. $w_m \ss \F$) and $\L^{(m)}$ (resp. $L^{(m)}_\F$). 

We stress that we have so far no guarantee that these local foliations (in particular the $\L^{(m)}$ or $L^{(m)}$ ones) patch together to give global ones in some tubular neighbourhood $N \ss M$ of $\La$, i.e. that they are integrable, and that we can thus use them in order to have a fibration of such a $N$.

\subsection{The Poincar\'e-Nekhoroshev map.}
 
Up to now we have just considered local geometry in $W_m$. However, we would like to have global results (fibration in tori); in order to do this, we will use a flow built out of the $X_i$ fields to go round the cycles of $\La$. 

We will use the fact that for any homotopy class in $\pi_1 (\La)$, there is a closed path $\ga$ on $\La$ with the given homotopy class and which can be realized as the orbit of some vector field obtained as a linear combination of the $X_i$'s. 
 
Thus, choose a point $m \in \La$. Given a non-contractible path $\^\ga$ on $\La$, this identifies a path $\ga$ such that $m \in \ga$ and which is in the same homotopy class as $\^\ga$; and a set of constants $c_i \in R$ such that $X_c = \sum c_i X_i$ has a periodic flow on $\La$, say with period one ($\^\ga$ not contractible guarantees $|{\bf c}| := \sum_i |c_i| \not= 0$), and $\gamma$ is the (closed) orbit under the flow of $X_c$ which passes through the reference point $m \in \La$. 

The whole construction can be pursued for any $m$ and for any homotopy class of $\gamma$, but for the moment we will think of $m$ and $\^\ga$ (i.e. of ${\bf c}$) as given. 
 
If we consider the time-one flow of points $p \in \S_m$ under $X_c$, this defines a map $\Theta$ from $\S_m$ to $M$, say $\Theta (p) := e^{X_c} p$. We know that, by construction, $\Theta (m) =
m$, while in general $p' := \Theta (p)$ is not only different from $p$, but can well fail to be in $\S_m$.

However, we know there will be a neighbourhood $D_m \subseteq \S_m \cap W_m$ such that $\Theta : D_m \to W_m$; let us restrict to consider $p \in D_m$, and call $\vartheta$ the restriction of $\Theta$ to $D_m$. For $p \in D_m$ we have $p' \in W_m$ and thus $p'$ will belong to a leaf $L_{p'}$ of the foliation $\L^{(m)}$. We can then define $p'' $ as the intersection of $L_{p'}$ with $\S_m$; we recall this is unique. 

In this way we have defined a map $\Psi_c^{(m)} : D_m \to \S_m$. We call this the {\bf Poincar\'e-Nekhoroshev map} (PN map). 
Note this is based at a point $m \in \La$ and depends on the constants ${\bf c} \in R^k$, i.e. on (the homotopy class of) the path $\gamma$. In the sequel we denote $\Psi_c^{(m)}$ simply by $\Psi$, at least while we think of $m$ and of $\gamma$ (and thus of ${\bf c}$) as fixed.
 
We note now that the map $\Psi_c^{(m)}$ can be seen the composition of two  flows, time-one flow along $X_c$ and the flow over a time $\delta$ (depending on $p'$, i.e. on $p$) along another vector field $X_\ell = \sum \ell_i X_i$: indeed, any two points on the same leaf of $\L^{(m)}$ can be connected by such a flow, by the very definition of $\L^{(m)}$. 

We write $p'' := \eta (p') = e^{\delta X_\ell} p'$, so that in the end the PN map $\Psi$ can be decomposed as $\Psi = \eta \circ \vartheta$. 

\sp
{\bf Remark 6.} As $\vartheta$ and $\eta$ are given by flows under $C^r$ vector fields, they are $C^r$ maps, and so is their composition. That is, recalling that $M$ and $F_i$ were also assumed to be $C^r$, and thus the $X_i$ are $C^r$ vector fields, we conclude that the PN map $\Psi$ is a $C^r$ map. \EOR

\section{Invariant tori}

\subsection{Fixed points of the Poincar\'e-Nekhoroshev map.}

Note that, by construction, $\Psi_c^{(m)} (m) = m$, i.e. any point $m \in \La$ is a fixed point for the PN map based at $m$, for any choice of $\ga$.

We can then wonder if there is any other point $p \in \S_m$ ($p \not= m$) for which $\Psi_c^{(m)} (p) = p$. In discussing this question, it will be useful to consider the linearization of $\Psi_c^{(m)}$ at the fixed point $m$; we denote this linear operator as $A_c^{(m)}$. For ease of notation, we will write $\Psi$ for $\Psi_c^{(m)}$, and $A$ for  $A_c^{(m)}$. It will be natural to consider the following 

\sp
{\bf Condition N.} {\it The spectrum of $A$ does not include 1.}

\sp
{\bf Remark 7.} Note that $1 \not\in {\rm Spec} (A)$ is a structurally stable relation. \EOR

In order to discuss fixed points of $\Psi$, we choose coordinates $(\phi, F , s)$ in $W_m$ (note that $w_m$ is obtained by choosing $\Fb = \beta_0$). We fix $\phi = \phi_0$, i.e. we consider points $p \in \S_m$ with $m$ given, and consider the map $ \Phi (p) := p - \Psi (p) $; fixed points of $\Psi$ corresponds to zeroes of $\Phi$. 

Using the coordinates $(\phi,F,s)$, we have that $p$ and $\Psi (p)$ have necessarily the same $\phi$ and $F$ coordinates: the invariance of $\phi$ is by construction (both points belong to $\S_m$), that of $F$ follows from the definition of the $X_i$ as hamiltonian flows of the $F_i$ and the commutativity condition. Thus we can see the $F$ as parameters, and have only to consider the $s$ coordinates: we can write, decomposing $\Psi$ and $\Phi$ in components according to the $s$ coordinates, 
$$ \Phi^i (s; F) \ := \ s^i - \Psi^i (s;F) \ . \eqno(1) $$

\sp
{\bf Lemma 1.} {\it Under our general hypotheses and assuming condition N is satisfied, (i) the zeroes of $\Phi$ are isolated in $\s_m$; (ii) in $\S_m$ there is a local $C^r$ $k$-parameter family of zeroes for $\Phi$.} 
\sp
{\bf Proof.} We know that $\Phi^i (0;\beta_0) = 0$; let us consider the Jacobian of this map in $m$. This is given by 
$$ (D \Phi)^i_j := {\pa \Phi^i \over \pa s^j } = \delta^i_j - {\pa \Psi^i \over \pa s^j} := E - A \ , \eqno(2) $$
where $E$ is the identity matrix and $A$ is the linearization of the map $\Psi$. It follows then that if the spectrum of $A$ does not contain 1, the Jacobian of $\Phi$ is not singular and therefore $m$ is an isolated zero of $\Phi^{(m)}_c$ in $\s_m$. 

Let us now consider variations of $\Phi$ in the $F$ directions as well:
$$ \d \phi^i (s,F) \ = \ \[ {\pa \phi^i \over \pa s^j} \, - \, {\pa \Psi^i \over \pa s^j} \] \, \d s^j \, - \, {\pa \Psi^i \over \pa F_a} \, \d F_a \ . \eqno(3) $$
Thus the kernel of $\d \Phi$ is identified by the condition
$$ \d s^i \ = \ [ E - A ]^{-1} \( {\pa \Psi^i \over \pa F_a} \) \d F_a \ ; \eqno(4) $$
the inverse on the right hand side exists under condition N; as already remarked, the fact that this is satisfied in $m$ guarantees it is also satisfied for $p \in \S_m$ sufficiently near to $m$, i.e. for $F - F(\beta_0)$ sufficiently small. The same consideration applies to $\d F_1 \wedge ... \d F_k \not=0$, which guarantees that the kernel of $\d \Phi$ is transversal to $\Fb^{-1} (\beta)$ for $|\beta - \beta_0 |$ sufficiently small.

That is, there is a local $k$-parameter family of points $(\phi_0,s,F)$, which can be written as $s = \mu (F) \in \S_m$, everywhere tangent to the kernel of $\d \Phi$. Moreover, by considering $(\pa s^i / \pa F_a )$ along this family, see (4) above, we have immediately that this family describes a $C^r$ manifold through $m$. \EOP

In slightly different words, we have shown that for any $\beta \in R^k$ with $|\beta - \beta_0|$ sufficiently small, there is a point $p = (\phi_0, s = \mu [F(\beta)],F(\beta) )$ which is a fixed point of the map $\Psi_c^{(m)}$.
If we parametrize this family by $\a^i = \beta^i - \beta_0^i$, we have that the set $\{ p = \mu (\a ; \phi_0) \}$ is a $C^r$ submanifold through $m$ in $\S_m$, with $\mu (0;\phi_0) = m \in \La$.

\subsection{Fibration in invariant tori.}

By the procedure described in the proof of Lemma 1, we can lift any point $m \in \La \ss \F $ to a fixed point of $\Psi_c^{(m)}$ [given by $p = \mu (\a ; \phi_0)$] on $\Fb^{-1} (\beta)$ for any $\beta$ sufficiently near to $\beta_0$, say $ |\a | := |\beta - \beta_0 | < \varepsilon$ (we denote the set of such $\a$ as $A_m$, and its radius as $a_m$). That is, we have a uniquely defined retract $\rho_\a (m)$ of $m$ on $\Fb^{-1} (\beta_0 + \a)$ for $\a \in A_m$. 

We denote by $B_{\cal Z}$ (resp. $B$) the infimum of $a_m$ through ${\cal Z} \ss \La$ (resp. through $\La$).

Let us now consider a fixed $\a \not= 0$ in $B$, and study how the $p = \mu (\a ; \phi )$ corresponding to different $\phi$ patch together. That is, we want to study the set $\rho_\a (\La )$. 

We will at this time assume that condition N holds (for the given homotopy class of the path $\ga$) for all points $m \in \La$; we will later see (in lemma 4) that actually the condition is either satisfied for all $m \in \La$, or for no $m \in \La$ at all. We will also denote $\La$ as $\La_0$.

\sp
{\bf Lemma 2.} {\it Let condition {\rm N} be satisfied at all points of a  contractible neighbourhood ${\cal Z} \ss \La$; then $\rho_\a ({\cal Z})$ is a $C^r$ local manifold for any $\a \in B_{\cal Z}$. If condition {\rm N} is satisfied for all $m \in \La$, then $\rho_\a (\La ) := \La_\a$ is a $C^r$ torus $\toro^k$, $C^r$-equivalent to $\La \equiv \La_{0}$.}

\sp
{\bf Proof.} The map $\mu (\a ; \phi) $ is obtained through an application of the implicit function theorem, thus it is differentiable of the same class as the map $\Psi_c^{(m)}$ with respect to the variables $\phi$; as noted in remark 6, $\Psi_c^{(m)}$ is a $C^r$ map, and thus the set $\M (\a_0 ) := \{ \mu (\a_0 , \phi ) \}$ is locally a $C^r$ manifold for any $\a_0 \in B_{\cal Z}$. 

Moreover, as $\mu (\a_0 ; \phi ) $ is unique for any given $\phi$, there is a global bijection between $\La_0$ and $\M (\a_0)$. To guarantee that $\M (\a_0 )$ is topologically the same as $\La \equiv \M (0)$, we just note that $\M (\a_0 )$ is a deformation retract of $\La$: this follows from $\lim_{\a \to 0} \mu (\a , \phi_0) = m (\phi_0 ) \in \La$, i.e. from the fact that $\rho_\a$ is continuous and $\rho_{0} (m) = m$ (in other words, $\M ( \varepsilon \a_0)$ realizes, for $\varepsilon \in [0,1]$, an homotopy between $\La_0 = \La$ and $\M (\a_0)$). Hence $\M (\a_0 ) $ and $\La$ are topologically equivalent.
 
The previous discussion shows that if they are topologically equivalent, then they are also $C^r$ equivalent (this can also be obtained directly by the fact that $\rho_\a (m)$ is $C^r$), and thus we conclude that $\M (\a )$ is a torus $\toro^k$, $C^r$-equivalent to $\La$. We will thus just write $\La_\a$ for $\M(\a )$; obviously $\La_{0} = \La$.

In this way, we have obtained that for any $\a \in B$ there is a torus $\La_\a$ of fixed points for the PN map $\Psi_c$. Note that we have not yet shown that this is actually an invariant manifold for the flows $X_i$. \EOP

\sp
{\bf Remark 8.} We have shown that in a neighbourhood $U \ss M$ of $\La$ there is a submanifold $N$ which is fibered in tori $\La_\a$; this is identified, in local coordinates, by $N = \{ (\phi , s, F) := s = \mu ( F ; \phi ) \}$. As $\d F_1 \wedge ... \wedge \d F_k \not= 0$ in a neighbourhood of $\La$, in a neighbourhood $\^N \sse N$ (corresponding to $\a \in \^B$) we can use the $\a$ as coordinates instead than the $F$, i.e. have that $\pi : \^N \to \^B$ is a fibration with tori $\toro^k$ as fibers, $\pi^{-1} (\a ) = \La_\a$. Obviously $\La_\a \ss \Fb^{-1} (\b_0 + \a )$, and actually $\La_\a = \^N \cap \Fb^{-1} (\b_0 + \a )$. \EOR

\sp
{\bf Lemma 3.} {\it The tori $\Lambda_\a$ are invariant under the $X_i$, i.e. $X_i : \Lambda_\a \to \T \La_\a$ for all $i = 1,...,k$ and for all $\a \in B$.}
\sp
{\bf Proof.} Let us consider a point $p \in \La_\a \cap \S_m$; we consider then a nearby point $p_\delta$ which is along a flow $X = \sum a_i X_i$, i.e. $p_\delta = e^{\delta X} p$.
Obviously if $p \in \Fb^{-1} (\b_0 + \a)$, then $p_\de \in \Fb^{-1} (\b_0 + \a)$ as well. We want to show that $p_\de$ is also in $\La_\a$, i.e. is a fixed point for $\Psi$. 

Let us denote by $m_\delta$ the point on $\La$ such that $p_\de \in \S_{m_\de}$, and consider how the $\Psi$ map acts on $p_\delta$: with a notation defined above, we have $(p_\de)' = \vartheta (p_\de) = e^{X_c} p_\de $, and ${p_\de}'' = L_{{p_\de}'} \cap \S_{m_\delta}$ (note we can use indifferently the $\L^{(m)}$ or the $\L^{(m_\delta )}$ foliations, as we are in $W_m \cap W_{m_\delta}$). 

Using $p_\delta = e^{\delta X} p$, we have  $ {p_\delta}' = e^{X_c} p_\de = e^{X_c} e^{\delta X} p $. Using now that fact that $X_c$ and $X$ necessarily commute (since the $X_i$ commute with each other), we have ${p_\delta}'  = e^{\delta X} e^{X_c} p = e^{\delta X} p' $. Thus, if $\delta$ is small enough -- that is, if $p', {p_\delta}'$ belong to $W_m \cap W_{m_\delta}$ -- then $p'$ and ${p_\delta}'$ belong to the same leaf, both in the $\L_c^{(m)}$ and in the $\L_c^{(m_\de)}$ foliations. 

Focusing on $\L^{(m_\delta )}$, we have in particular $[L^{(m_\de)}_0]_{{p_\de}'} \cap \S_m = p'' = p$. But we know that $p_\de = e^{\de X} p$; and thus it belongs to the same local leaf as $p = p''$; obviously $e^{\de X} p$ does also belong, by definition, to $\S_{m_\de}$, and thus ${p_\de}'' := [L^{m_\de}_0]_{{p_\de}'} \cap \S_{m_\de} = p_\de$.

This shows that for $|p_\de - p |$ sufficiently small, $e^{\de X} $ takes points in $\La_\a$ to points in $\La_\a$, for any $X = \sum a_i X_i$, i.e. that any such $X$ is tangent to $\La_\a$. Thus, $X_i : \La_\a \to \T \La_\a$. \EOP

\sp
{\bf Lemma 4.} {\it The $C^r$ manifold $N$ obtained as the union of $\Lambda_\a$ for $\a \in B$ is symplectic and fibered in isotropic tori $\toro^k$.}
\sp
{\bf Proof.} It is clear that $N$ is $C^r$ and that it is fibered in tori $\toro^k \equiv \Lambda_\a$. The tori $\Lambda^\a$ are integral manifolds of the hamiltonian vector field generated by the functions $F_i$, and lie in common level manifolds of the $F_i$; thus we can choose a basis of angular variables $\phi_i$ along $\Lambda_\a$ such that $\omega (\pa / \pa F_i , \pa / \pa \phi_j ) = \delta_{ij}$. This shows at once that the $\Lambda_\a$ are isotropic and that the restriction $\om_N$ of the symplectic form $\om$ defined on $M$ to the submanifold $N \ss M$ is non degenerate. As $\d \om = 0$ implies $\d \om_N = 0$, the proof is complete. \EOP

This concludes the proof of the Poincar\'e-Lyapounov-Nekhoroshev theorem, and nearly (but not fully) concludes our discussion of it. Indeed, we should still examine the dependence of our construction and proof on the arbitrary choices on which it is based.

\section{Dependence on arbitrary choices}

We have so far considered a given path $\^\ga$ on $\La$, i.e. a vector field $X_c$, and a reference point $m \in \La$; actually this reference point was then somehow abandoned, as we assumed (see sect.2.2) that the condition $1 \not\in {\rm spec} (A_c^{(m)} )$ is satisfied at all $m \in \La$. We want to discuss the relevance of these choices. 

We will obtain in lemma 5 that the choice of $m \in \Lambda$ is immaterial, i.e. that once we have chosen a $\^\ga$ (which defines the PN map), condition N is satisfied (or violated) at all points of $\La$ at once, so that we could indeed work only at a reference point. We will also discuss in lemma 6 how to check easily if condition $N$ is satisfied (for a given homotopy class of $\^\ga$). 

On the other side, as we discuss in lemma 7, the choice of the path $\^\ga$ affects condition $N$ being satisfied or otherwise.

\subsection{Condition N and the reference point on the torus}

We will consider loops along fundamental cycles of the torus $\La$; these can be realized as orbits of vector fields in the linear span of the $X_i$, say as time-one flow of vector fields $Y_i = \sum \nu_i^j X_j$. 
Having built the fibration of $N$ in tori $\La_\b$, we can extend the $Y_i : \La \to \T \La$ to vector fields $Z_i : N \to \T N$ which are along the fibers of $N$, i.e. such that $Z_i : \La_\b \to \T \La_\b$ for all $\b \in B$, and which act in the same way on all the $\La_\b$, i.e. such that the flow along $Z_i$ commutes with the action of the retract $\rho_m$; this follows from the fact the $\rho_m$ is an equivariant retract under the (torus) action of the $X_i$. 

It can be helpful to remark that the $Z_i$'s can be written as $Z_i = \zeta_i^j (\b) X_j$, and that the coefficients $\zeta_i^j (\b)$ are therefore constant on each $\La_\b$.
This means that given points $m_a ,m_b \in \La$ and a vector field $Z = \sum \a_i Z_i$ such that $m_b = e^Z m_a$, we can choose $\S_{m_b} = e^Z \S_{m_a}$.

\sp
{\bf Lemma 5.} {\it If the condition $1 \not\in {\rm Spec} (A_c^{(m_a)})$ is satisfied (respectly, violated) at a point $m_a \in \La$, then $1 \not\in {\rm Spec} (A_c^{(m_b)})$ is also satisfied (resp. violated) at all points $m_b \in \La$.}
\sp
{\bf Proof.} We can pass from any point $m_a \in \La$ to any other point $m_b \in \La$ by a translation along $\La$; we realize this by a $Z_i$ action, i.e. $ m_b = e^Z m_a$ with $Z = \sum \a_i Z_i$. We write the PN maps based at $m_a$ and at $m_b$ as $\^p_a := \Psi_c^{(m_a)} (p) = e^{Y_a} e^{X_c} p$, and the like for $m_b$. However, given a point $p_b \in \S_{m_b}$, there will be a point $p_a \in \S_{m_a}$ such that $p_b = e^Z p_a$. Thus we have, using commutativity of flows at various stages,
$$ \begin{array}{rl}
\^p_b := & \Psi_c^{(m_b)} (p_b) \ = \ e^{Y_b} e^{X_c} e^Z p_a \ = \ e^Z e^{Y_b} e^{X_c} p_a \ = \\ 
 & \ = \ (e^Z e^{Y_b} e^{- Y_a} ) \Psi_c^{(m_a)} (p_a) \ = \ ( e^{Y_b} e^{- Y_a} ) e^Z \Psi_c^{(m_a)} (p_a) \ = \\
 & \ = \ ( e^{Y_b} e^{- Y_a} ) e^Z \^p_{m_a} \end{array} \eqno(5) $$
This means that $\^p_b$ and $e^Z \^p_a$ belong to the same leaf in the 
$ \L^{(m_b)}_c$ foliation; however, we know that since $\^p_a \in \S_{m_a}$, necessarily $e^Z \^p_a \in \S_{m_b}$, i.e. we have shown that $\^p_b = e^Z \^p_a$. In other words,
$$ e^Z \Psi_c^{(m)} (p) \ = \ \Psi_c^{(e^Z m)} (e^Z p) \ . \eqno(6) $$
This also shows that $e^Z$ conjugates $A_c^{(m_a)}$ and $A_c^{(m_b)}$, i.e. that 
$$ A_c^{(e^Z m)} \ = \ e^Z \, A_c^{(m)} \, e^{-Z} \ ; \eqno(7) $$
the assertion of the lemma follows immediately from this relation. \EOP

We have thus shown that in order to ensure condition N is satisfied (for a given choice of the loop $\^\ga$) at all points of $\La$, it is enough to check it is satisfied at a given reference point $m \in \La$.

\sp
{\bf Lemma 6.} {\it The spectrum of the linear map $A_c^{(m)}$ coincides with the spectrum of the linearized action of the monodromy operator $\exp [ X_c ]$ on directions transversal to $\Lambda_0$ in ${\bf F}^{-1} (\b )$.}

\sp {\bf Proof.}
We consider coordinates $(\a , \phi ; s)$ as above. Recall that the PN map $\Psi$ can be written as $\Psi = \eta \cdot 
\vartheta$; however, if we use the coordinates mentioned above, it should be noted that, by definition, the $\eta$ does act 
only on the $\phi$ coordinates, and not on the other ones. We write $\vartheta (\a , \phi ; s) = (\a , \~\phi ; \~s )$; 
the action of $\eta$ is then (by definition of our coordinate system) 
$\eta (\a , \~\phi ; \~s) = (\a , \phi , \~s ) := \Psi (\a , \phi ; s)$.

Let us write the linearization $B$ of $\Psi$ (without restriction to the $\s_m$ manifold) in the $(\a , \phi ; s )$ 
block form: by the previous formula, 
$$ B \ = \ \pmatrix{I & 0 & 0 \cr 0 & I & 0 \cr 0 & M  & B_{(s)} \cr} $$
where $M = (\pa \~s^i / \pa \phi_k )$ is some rectangular matrix we are not interested in, and $B_{(s)}$ is the projection to 
the subspace spanned by the $s$ variables of the linearization of the $\vartheta$ map, i.e. is determined by the partial 
derivatives $\pa \~s^i / \pa s^j$. 
Obviously the spectrum of $B$ consists of the eigenvalue $\lambda_0 =1$ with multeplicity $2k$, corresponding to the first 
two blocks -- each of them having dimension $k$ -- in the $B$ matrix, and of the spectrum of the $B_{(s)}$ matrix. 

However, as pointed out above, $B_{(s)}$ is nothing else than the projection to the $s$ subspace of the linearization of the 
$\exp [X_c]$ operator (at $\a, \phi)$ fixed. Hence the lemma. $\ \triangle$

\sp
{\bf Corollary.} {\it Condition N is satisfied if and only if the spectrum of $A_{(s)}$ does not contain the unity.}

\subsection{Condition N and the reference path on the torus}

Our proof of Nekhoroshev's theorem was also based on a specific choice of $X_c$; we recall that given any nontrivial closed path $\ga$ in $\La$, one can find a closed path ${\widetilde \ga}$ homotopic to $\ga$ which is the orbit under a vector field of the form $X = \sum c_i X_i$ (say with period one when $|c| \not= 0$), and this is precisely how we associate to $\ga$ the vector field $X_c$ (see sect.1.4). Thus, we have based our proof on a specific choice of the homotopy class of $\ga$; we want to show how the holding or otherwise of condition N depends on our choice of the homotopically nontrivial path $\ga$. 

Consider a loop $\ga'$, not homotopically equivalent to $\ga$; there is then a ${\widetilde \ga'}$, homotopically equivalent to $\ga'$,  which is the orbit under a vector field of the form $X' = \sum {c'}_i X_i = X_{c'}$ (say with period one when $|c'| \not= 0$). We can repeat our construction based on this vector field $X_{c'}$ rather than on $X_c$; in particular we denote the PN map based on $X_{c'}$ (and with reference point $m \in \La$) as $A_{c'}^{(m)}$.

The fact that if condition N is satisfied for $A_{c'}^{(m)}$ -- so that we can actually perform our construction -- we would obtain the same tori as when using the PN map $A_{c}^{(m)}$ based on $\ga$, is obvious: they are invariant tori for all fields $X$ in the span of the $X_i$. 

We wonder then if the holding or otherwise of condition $N$ (say at a given point $m \in \La$, see lemma 4 above) can actually depend on  the (nontrivial, otherwise the PN map is itself trivial) homotopy class of $\ga$, i.e. if condition $N$ can be violated for $\ga$ but satisfied for a homotopically nonequivalent loop $\ga'$.
It turns out that this is the case.

\sp
{\bf Lemma 7.} {\it Consider $m \in \Lambda_0$ and two paths $\ga_1$ and $\ga_2$, homotopically non-equivalent, through $m$. Let $\Psi_1 , \Psi_2$ be the PN maps at $m$ based on the paths $\ga_1$ and $\ga_2$, and let $A,B$ be their linearizations. In general, ${\rm spec} (A) \not= {\rm spec} (B)$.}
\sp
{\bf Proof.} We can always pass from one homotopy class to another by adding a loop along fundamental cycles of the torus $\La$; these correspond to time-one flow of the vector fields $Z_i$ considered above. Thus, any two homotopy classes of paths in $\La$ have representatives $\ga$, $\ga'$ through $m$ which represent orbits of vector fields $X_c$, $X_{c'}$ and there is a vector field $Z = \sum_i n_i Z_i$ with integer coefficients $n_i \in {\bf Z}$ such that $X_{c'} = Z + X_c$.
 
Then if we choose a curve $\^\ga'$ non homotopic to $\^\ga$, the flow on $\La$ of the vector field $X_{c'}$ corresponding to the curve $\ga'$ has orbits which are homotopic to the orbits of $X_{c''} = (Z + X_c)$, with $Z = \sum n_i Z_i$. However, we cannot consider such a flow in the construction of the PN map, as it is not a linear combination (with constant coefficients) of the $X_i$, but is just in the module over smooth functions of the $F_i$ generated by the $X_i$. We have thus to consider instead the vector field $Y = \sum_i n_i Y_i$, where the coefficients $n_i$ are the same as in $Z = \sum_i n_i Z_i$, and the $Y_i$ generate the cycles of $\Lambda_0$ (they have been defined above). 

The map $\Psi_{c''}^{(m)} (p) $ can be written, as usual, as
$\Psi_{c''}^{(m)} (p) = L_{e^{Y+X_c}p} \cap \S_m = L_{e^Y e^{X_c} p} \cap \S_m$. Call now $X_\de$ the vector field (in the span of the $X_i$) such that $p'' := L_{e^{Y+X_c}p} \cap \S_m = e^{X_\de} p' $. We thus have, using commutativity of the $X_i$ and of linear combinations thereof,
$$ \Psi_{c''}^{(m)} (p) \ = \ e^{X_\de} \, e^{Y + X_c} \, p \ = \ e^Y \[ e^X_\de  e^{X_c} p \] \ \equiv \ e^Y \[ \Psi_c^{(m)} (p) \] \ . \eqno(8) $$

We write $A$ for the linearization of $\Psi_c^{(m)}$ and $B$ for the linearization of $\Psi_{c''}^{(m)}$; let $DY$ be the linearization of the vector field $Y$ (in $\Psi_c^{(m)} (p)$). Then (8) reads $B = A + DY$, and obviously the spectrum of $B$ is in general different from the spectrum of $A$. \EOP

\sp
{\bf Remark 9.} This lemma is relevant in applications; indeed, given an invariant torus $\Lambda$, we can work with a given path $\ga_1$ and establish it can be continued to a family on invariant tori $\Lambda_\eps$ for $\eps < \eps_1$ by studying the spectrum of $A$ (we are considering the same notation as in the lemma). When we reach the border of this region, i.e. a torus $\La_1$ (corresponding to $\eps = \eps_1$) for which the monodromy operator associated to the cycle $\ga_1$ becomes singular, however, it is still possible that passing to consider a different path $\ga_2$ we can guarantee that the family of invariant tori can be continued up to some $\eps_2 > \eps_1$. In other words, the family can be continued provided at least one of the monodromy operators associated to the fundamental cycles $\ga_\a$ of the tori is nonsingular (has spectrum satisfying condition N). \EOR

\sp
{\bf Remark 10.} We stress that when we consider the problem of constructing action-angle variables on the submanifold $N$ fibered by the invariant tori $\La_\b$, we can in general obtain global action-angle coordinates only on each set $\b_i^- < \b_i < \b_i^+$ for which all the monodromy operators are nonsingular; the obstruction to having global action-angle coordinates on $N$ is associated to monodromy, and is the same as the one in the integrable case [indeed the system is integrable on $(N,\om_N)$]; see e.g. \cite{Arn1,Dui} or the simple discussion in \cite{Cor}. \EOR

\section{Symmetrically perturbed oscillators.}

Let us consider an Hamiltonian in $r$ degrees of freedom with an equilibrium point in the origin. We write it as $H = H_0 + G$, where $H_0$ is the quadratic part and $G$ contains higher order terms only. We write the quadratic part as
$$ H_0 \ = \ \sum_{i=1}^n \ \om_i \ {p_i^2 + q_i^2 \over 2} \ , \eqno(9) $$
i.e. as a collection of oscillators; we assume that the $\om_i$ are nonzero and different from each other. 

We assume now that $I_k := (1/2) (p_k^2 + q_k^2 )$ are constants of motion for $k=1,...,s-1$; in other words, we suppose that $H$ admits $s$ constants of motion in involution -- one of them being $H$ itself -- and that they are explicitely known. We write $\phi_k$ for the action variable conjugated to the action variables $I_k$. 

Note that this case is met when we have a quadratic hamiltonian $H_0$ perturbed by nonlinear terms, with the perturbation being symmetric under the abelian group $U(1) \times ... \times U(1)$ generated by the vector fields 
$$ X_k \ := \ {\pa \over \pa \phi_k } \ = \ p_k \, {\pa \over \pa q_k} \,  - \, q_k \, {\pa \over \pa p_k} \ \ \ , \ k=1,...,s-1 . $$

We also write, for $j = 1,...,r:=n+1-s$, 
$$ p_{s-1+j} \ = \ A_{j} \, \cos (\psi_j) \ \ , \ \ q_{s-1+j} \ = \  A_{j} \, \sin (\psi_j) \ . $$ 
We also rewrite the frequencies corresponding to the $\psi_j$ variables as $\nu_j$ in order to emphasize the difference with the $\phi$ variables. With this notation, the hamiltonian is rewritten as
$$ H \ = \ \sum_{k=1}^{s-1} \, \om_k I_k \ + \ \sum_{j=1}^r \, \nu_j A_j \ + \ G (I;A,\psi) \ .  \eqno(10)$$
We assume that $G$ is such that 
$$ G(I;0,\psi) \ = \ {\pa G \over  \pa I_k } (I;0,\psi) \ = \ 0 \ . \eqno(11)$$

We will consider the family of commuting integrals $\{ F_1 , ... , F_s \}$ given by $F_k = I_k$ for $k < s$, and $F_s = H$. With this choice, and writing $\pa_k \equiv \pa / \pa \phi_k$, the commuting vector fields are given by 
$$ X_k = \pa_k \ (k < s) \ \ , \ \ X_s \equiv X_H = \sum_k \om_k \pa_k + \sum_j \nu_j (\pa / \pa \psi_j ) + X_G \ . $$

The tori $I_k = c_k$ and $A_1 = ... = A_r = 0$ are obviously 
invariant. We denote by $\Pi$ the operator of projection of vector fields to these tori. 

The most general linear combination of the $X_i$ is $X_c = \sum_{i=1}^{s-1} \a_i X_i + \b X_s$; the projection $Y_c := \Pi X_c$ of these to the invariant tori is
$$ Y_c \ = \ \sum_{k=1}^{s-1} \, \a_k \, \pa_k \ + \ \b \[ \sum_{k=1}^s \, \om_k \, \pa_k \] \ = \ \sum_{k=1}^{s-1} \( \a_k + \b \om_k \) \, \pa_k \ + \ \b \om_s \pa_s \ . $$

The condition for $Y_c$ to have closed orbits $\ga$ of period $T = 2 \pi \tau$ and winding numbers $n_k$ with respect to the cycles of the torus, described by the variables $\phi_k$, is that
$$ Y_c \ = \ \sum_{k=1}^s \ \tau \, n_k \, \pa_k \ ; $$
and therefore we must require to have
$$ \cases{
\a_k + \b \om_k \ = \ \tau n_k & for $k=1,...,s-1$ \cr 
\b \om_s = \tau n_s \ ; & \cr} \eqno(12) $$
note that we can (and will) set $\tau = 1$ by rescaling the $\a_k$ and $\b$.

If $n_s = 0$ (i.e. if $X_H$ does not enter in the vector field $X_c$), then we have $\b = 0$ and $\a_k = n_k$; however we will see in a moment that condition N can never be satisfied in this case.

We will assume $n_s \not= 0$, so that the solution to (12) is
$$ \b = {n_s \over \om_s} \ ; \ \a_k = {1 \over \om_s } \[ n_k \om_s - n_s \om_k \] \ . \eqno(13) $$

We can thus, for any set of integers $\{ n_1 , ... , n_s \}$, determine the vector field $X_c$ entering in the construction of the Poincar\'e-Nekhoroshev map. 

In order to check if condition N is satisfied or otherwise, it suffices to compute the Floquet multipliers $f_j$ in directions 
transversal to invariant tori. These are related to rotation numbers for the $\psi_j$ angles, and writing down $X_c$ 
we have immediately that $f_j = \exp [\b \nu_j ]$ (note that $\b = 0$ implies $f_j = 1$ for all $j$, i.e. condition N is not satisfied); thus, using (13), condition N is equivalent to 
$$ n_s \ {\nu_j \over \om_s } \ \not\in \ {\bf Z} \ \ \ \ \forall j=1,..,r \ . \eqno(14) $$

\bigskip

Note that -- with $\mu_k = \a_k (1 - \delta_{ks})$ and $g_k = (\pa G / \pa I_k)$ for ease of writing -- the complete expression of $X_c$ would be $$ \begin{array}{rl}
X_c \ =& \ \sum_{k=1}^{s} \[ \mu_k + \b \om_k + \b g_k \] \, {\pa \over \pa \phi_k} \ + \\ 
 & + \ \b \sum_{j=1}^r \[  \[ \nu_j p_j + (\pa G / \pa p_i ) \] {\pa \over \pa q_j} - \[ \nu_j q_j + (\pa G / \pa q_j ) \] {\pa \over \pa q_j} \] \ . \end{array} $$

If we drop (11) and look at one of the invariant tori $I_k = c_k , A_j = 0$, then (12) are transformed into 
$$ \mu_k + \b ( \om_k + g_k ) \ = \ \tau n_k \ , \eqno(12') $$
with solution
$$ \b = n_s / (\om_s + g_s ) \ \ ; \ \ \a_k = n_k (\om_s + g_s ) - n_s (\om_k + g_k ) \ . \eqno(13') $$
The Floquet multipliers are still $f_j = \exp[ \b \nu_j ]$, and thus condition N reads now
$$ n_s \ {\nu_j \over \om_s + g_s } \ \not\in \ {\bf Z} \ \ \ \ \ \forall j=1,..,r \ . \eqno(14')$$

\section{Perturbed oscillators II.}

We consider a variation of the setting of the previous section, retaining the notation introduced there. That is, we consider  perturbations of a $n$-dimensional harmonic oscillator with $s-1$ additional constants of motion $F_i$ beyond the hamiltonian $H$ (we will also write $F_s = H$), but now we do {\it not } assume that the additional integrals of motion are known.

We write again $H = H_0 + G$ with $H_0$ as in (9), and we assume again that the $\om_i$ are nonzero and different from each other. 

It follows from $\{ H , F_i \} = 0$ that, expanding $F_i$ in a power series, the $F_i$ have a quadratic part $F_i^{(0)}$ which necessarily commutes with $H_0$: this means that necessarily 
$$ F_i^{(0)} = \sum_{k=1}^n \La_{ik} I_k \ \ \ , \ \ i=1,...,s-1 \eqno(15) $$
where $I_k = (p_k^2 + q_k)^2/2$. We will use the notation $p_k = I_k \cos (\phi_k ) $, $q_k = I_k \sin (\phi_k )$.

\sp
{\bf Remark 11.} This case corresponds to $\La$ a reducible torus for the functions $(F_1, ... , F_s)$; conditions ensuring this are discussed e.g. in \cite{Kuk} and references therein. Note that here we suppose the $F_i$ are not explicitely known. \EOR

The linear hamiltonian fields associated to the $F_i^{(0)}$ are therefore, with this notation and writing $\pa_k := (\pa / \pa \phi_k)$, 
$$ X_i \ = \ \sum_{k=1}^n \ \La_{ik} \pa_k \ \ \ (i=1,...,s) \ . $$
We stress that this is just the linear part of the vector fields tangent to the invariant tori; to obtain full fields we should add a nonlinear part. By rescaling variables $(p,q)$ we will have that nonlinear terms correspond to a correction $O( \eps)$, so that our analysis corresponds to the case $\eps = 0$.

We will, for ease of discussion, write the matrix $\La$ in block form as 
$$ \La \ = \ \pmatrix{ A & | & B \cr } $$
where $A$ is a $(s \times s)$ matrix, and $B$ is a $(s \times r)$ one, with $r := (n - s)$.

Suppose that the ordering of the $I_k$ is such that $\{ \pa_1 , ... , \pa_s \}$ are tangent to invariant tori to order $\eps$ (if not, we can always reduce to this case by means of linear transformations).
Thus the projection $Y_c$ of a general vector field $X_c = \sum_{i=1}^s \a_i X_i$  to invariant tori is just the projection onto $\{ \pa_1 , ... , \pa_s \}$. Hence,
$$ Y_c \ = \ \sum_{i=1}^s \ \sum_{k=1}^s \ \a_i \La_{ik} \pa_k \ \equiv \ \sum_{i=1}^s \ \sum_{k=1}^s \ \a_i A_{ik} \pa_k \ .  \eqno(16) $$
When we require that this has closed orbits with winding number $n_i$ around the cycles of the torus given by $\phi_i$, we are requiring $Y_c = \sum_{k=1}^s \tau n_k \pa_k $ for some relatively prime integers $n_k$; as in previous example, we will set $\tau = 1$, which can always be obtained by rescaling the $\a_i$. Thus we are asking
$ \a_i A_{ik} = n_k$, i.e. $A^T \a = n$, and we get 
$$ \a_i \ = \ P_{ij} \, n_j \ \ ; \ \ P = (A^T)^{-1} \ . $$

Let us now look at the flow on the other angles $\phi_{s+1} , ... , \phi_{n}$; we will for ease of notation write $\psi_j := \phi_{s+j}$, where $j=1,...,r:=n-s$. Obviously we have
$$ {\dot \psi_j} \ = \ Q_j \ = \ \a_i \La_{i,j+s} \ := \ (B^T)_{ji} \a_i \ = \ B^T_{ji} P_{ik} n_k \ , $$
and condition N is satisfied (for this choice of the $n_k$) if $Q_j \not\in {\bf Z}$ for all $j=1,...,r$. In other words, it is satisfied if for all choices of $m = (m_1,...,m_r) \in {\bf Z}^r$,  
$$ (B^T)_{i j} \, P_{j k} \, n_k \ \not=  m_i \ . \eqno(17) $$

\bigskip

It is interesting to note that this condition can be expressed in terms of the nonvanishing of a linear combination (with integer coefficients) of determinants, as we briefly discuss now.

As we have to deal with the transpose of the matrix $A$, we will denote $A^T$ by $\Om$. We will denote by $\Om^*(k;j)$ the matrix obtained from $\Om$ by substituting the $k$-th row with the $j$-th row of $B^T$.

We then recall that, by Cramer's theorem, $P := \Om^{-1}$ is obtained as $P_{ij} = (| \Om |)^{-1} \bar{\Om}^T$, where we denote by $\bar{M}$ the matrix of algebraic complements of a matrix $M$, i.e. $\bar{M}_{ij}$ is the algebraic complement of $m_{ij}$ in $M$. Here and below $|M|$ denotes the determinant of the matrix $M$; from elementary linear algebra we know that $\sum_\ell m_{j\ell} \bar{M}_{k \ell} = \delta_{jk} |M|$. Note that by construction $[\Om^* (k,j)]_{k \ell} = (B^T)_{j \ell}$ and, again by construction, the algebraic complement of $[\Om^* (k,j)]_{k \ell}$ in $\Om^* (k,j)$ is the same as the algebraic complement of $(A^T)_{k \ell}$ in $A^T$. 

It is thus clear, using these facts, that 
$$ \sum_{i=1}^s \ (B^T)_{ji} \, P_{ik} \ = \ { | \Om^* (k;j) | \over |\Om | } \ . \eqno(18) $$ 

Therefore, the condition that $Q_j \not\in {\bf Z}$ can be reformulated as the condition that 
$$ \sum_{k=1}^s \ n_k \, | \Om^*(k;j) | \ \not= \ m \ |\Om | \ \ \forall m \in {\bf Z} \ . \eqno(19) $$

We will now consider some simple explicit examples, in order to show that this criterion is easily checked. 

More complex applications will be considered elsewhere: in \cite{BG} this is applied to elliptic tori and the reduced three-body problem, in \cite{BV} it is applied to study the existence of breathers in infinite chains of coupled nonlinear oscillators.

\subsection{Example 1.}

Let us consider the simplest nontrivial case, i.e. $n=3$, $s=2$. We will write
$$ X_1 = \om_1 \pa_1 + \om_2 \pa_2 + \om_3 \pa_3 \ \ , \ \ 
X_2 = \mu_1 \pa_1 + \mu_2 \pa_2 + \mu_3 \pa_3 \ , $$
and $X_c = \a X_1 + \b X_2$, so that $Y_c = (\a \om_1 + \b \mu_1 ) \pa_1 + (\a \om_2 + \b \mu_2 ) \pa_2$. 
In this case,
$$ \La \ = \ \pmatrix{\om_1 & \om_2 & \om_3 \cr \mu_1 & \mu_2 & \mu_3 \cr} \ \ ; \ \ \Om \ = \ \pmatrix{\om_1 & \mu_1 \cr \om_2 & \mu_2 \cr} \ \ , \ \ B^T \ = \ \pmatrix{\om_3 & \mu_3 \cr} \ . $$
We have 
$$ \Om^{-1} \ = \ {1 \over | \Om |} \ \pmatrix{\mu_2 & - \mu_1 \cr - \om_2 & \om_1 \cr} $$
and hence
$$ Q_1 = B^T \, P \, \pmatrix{n_1 \cr n_2 \cr } \, := \, B^T \, \pmatrix{\a \cr \b \cr} \, = \, {1 \over |\Om |} \left[ n_1 (\mu_2 \om_3 - \om_2 \mu_3 )  +  n_2 (\om_1 \mu_3 - \mu_1 \om_3 ) \right] . $$
Thus $Q_1 \not\in {\bf Z}$ is rewritten as 
$$ n_1 \left| \matrix{\om_3 & \mu_3 \cr \om_2 & \mu_2 \cr} \right|
\ + \ n_2 \left| \matrix{\om_1 & \mu_1 \cr \om_3 & \mu_3 \cr} \right| \ \not= \ K \left| \matrix{\om_1 & \mu_1 \cr \om_2 & \mu_2 \cr} \right| \ \ \forall K \in {\bf Z} \ . $$

\subsection{Example 2.}

Let us now see the simplest case with $r \not= 1$, i.e. the one corresponding to $n=4$, $s=2$. We write
$$ X_1 = \om_1 \pa_1 + ... + \om_4 \pa_4 \ \ , \ \ X_2 = \mu_1 \pa_1 + ... + \mu_4 \pa_4 $$
The matrices $\Om$ and $P = \Om^{-1}$ are thus the same as above; now
$$ B^T \ = \ \pmatrix{\om_3 & \mu_3 \cr \om_4 & \mu_4 \cr} $$
and it is immediate to check that 
$$ B^T \, P \ = \ {1 \over |\Om | } \ \pmatrix{
\left| \matrix{\om_3 & \mu_3 \cr \om_2 & \mu_2 \cr} \right| & 
\left| \matrix{\om_1 & \mu_1 \cr \om_3 & \mu_3 \cr} \right| \cr 
 ~ & ~ \cr  ~ & ~ \cr
\left| \matrix{\om_4 & \mu_4 \cr \om_2 & \mu_2 \cr} \right| & 
\left| \matrix{\om_1 & \mu_1 \cr \om_4 & \mu_4 \cr} \right| \cr } $$

Thus, $Q_3$ and $Q_4$ are solutions of the equations
$$ 
n_1 \, \left| \matrix{\om_3 & \mu_3 \cr \om_2 & \mu_2 \cr} \right| \ + \ n_2 \, \left| \matrix{\om_1 & \mu_1 \cr \om_3 & \mu_3 \cr} \right| \ = \ Q_3 \, \left| \matrix{\om_1 & \mu_1 \cr \om_2 & \mu_2 \cr} \right| $$
$$ n_1 \, \left| \matrix{\om_4 & \mu_4 \cr \om_2 & \mu_2 \cr} \right| \ + \ n_2 \left| \matrix{\om_1 & \mu_1 \cr \om_4 & \mu_4 \cr} \right| \ = \ Q_4 \, \left| \matrix{\om_1 & \mu_1 \cr \om_2 & \mu_2 \cr} \right| $$
which of course are rewritten, in the notation introduced above, as 
$$ \cases{
n_1 \, | \Om^*(1;3) | \ + \ n_2 \, |\Om^* (2;3) | \ = \ Q_3 \, |\Om | & \cr 
n_1 \, | \Om^*(1;4) | \ + \ n_2 \, |\Om^* (2;4) | \ = \ Q_4 \, |\Om | & \cr } $$
These are readily solved, and we get
$$ \begin{array}{ll}
Q_3 & \ = \  [n_1 (\om_3 \mu_2 - \om_2 \mu_3 ) \ + \ n_2 (\om_1 \mu_3 - \om_3 \mu_1 ] \ / \ (\om_1 \mu_2 - \om_2 \mu_1 ) \\
Q_4 & \ = \ [n_1 (\om_4 \mu_2 - \om_2 \mu_4 ) \ + \ n_2 (\om_1 \mu_4 - \om_4 \mu_1 ] \ / \  (\om_1 \mu_2 - \om_2 \mu_1 ) \end{array} $$

The condition $Q_i \not\in {\bf Z}$, which can be readily checked by the above explicit formula, is also rewritten as 
$$ n_1 \, | \Om^*(1;i) | \ + \ n_2 \, |\Om^* (2;i) | \ \not= \ K \, |\Om | \kern 1 cm \forall K \in {\bf Z} \ , \ \forall i = 3,4 \ . $$

\vfill\eject

\section*{Appendix. The non-hamiltonian case.}
\def\G{{\cal G}}

We note that several parts of Nekhoroshev's results also extend to the non-hamiltonian case, employing essentially the same construction (i.e. the Poincar\'e-Nekhoroshev map). 

Indeed the role of the functions $F_i$ was essentially to identify $\La_\b$ with $N \cap \Fb^{-1}$, and later on to guarantee by construction the simplecticity of $N$; the rest of the discussion was rather based on the vector fields $X_i$. 

Thus we can reformulate the construction and the result in the case where there are $k$ commuting vector fields $X_i$ and an invariant torus $\La$. Needless to say, in this case we obtain only the result on fibration in manifolds diffeomorphic to $\La$ (i.e. in tori) but we cannot consider action-angle variables nor the restriction to level manifolds of the $F_i$. 

For non-hamiltonian vector fields, it would in many ways be natural to consider the case where $M$ is equipped with a metric and thus $\T M$ is equipped with a scalar product. 
In this case one can consider the tangent subbundle $\T \Lambda \ss \T M$ and the normal bundle $N \Lambda \ss \T_\La M$ to $\Lambda$ as a natural building ground for the two foliations used to define the Poincar\'e-Nekhoroshev map. 

Let us go quickly through the non-hamiltonian version of the construction considered in the main body of the paper; for ease of writing, we will say "smooth" to mean "$C^r$ smooth". We consider a smooth real riemannian manifold $(M,g)$ of dimension $n$, and $k$ smooth vector fields $X_i$, almost everywhere independent on $M$ and mutually commuting, $[X_i , X_j ] = 0$. We denote by $\G$ the abelian  Lie algebra spanned by the $X_i$. 

By passing to the one forms $\xi_i$ dual to the vector fields $X_i$ (i.e. satisfying $X_i \interno \xi_j = \delta_{ij}$), the linear independence of the $X_i$ reads 
$ \eta := \xi_1 \w ... \w \xi_k \not= 0$.

We assume there is a torus $\La = \toro^k$ invariant under $\G$, i.e. $X_i : \La \to \T \La$ for all $i=1,...,k$, and such that $\eta \not= 0 $ on $\La$. 

Consider through each point $m \in \Lambda$ the geodetic local manifold $\S_m$ orthogonal to $\La$ in $m = \La \cap \S_m$. We consider geodetic coordinates on these. 
In this way we can consider a tubular neighbourhood $U$ of $\La$ as a fiber bundle $\pi : U \to \La$ over $\La$ with fiber $\pi^{-1} (m) = \S_m$. 

Let us now consider a neighbourhood ${\cal Z}_0$ of $m$ in $\La$, and a neighbourhood ${\cal U}_0$ of $m$ in $U$, such that $\pi ({\cal U}_0) = {\cal Z}_0$ and $X_i : {\cal U}_0 \to \T {\cal U}_0$ (for all $i$); thus $\pi: {\cal U}_0 \to {\cal Z}_0$ is a local bundle.
We equip this local bundle with a connection $\Gamma$ generated by the vector fields $X_i$, i.e. given by a distribution of horizontal planes which is just the space spanned by the $X_i$; note that $\eta \not= 0$ on $\Lambda$ and smoothness of the $X_i$ guarantee that for ${\cal U}_0$ small enough this is indeed a distribution of horizontal $k$-planes, and thus defines actually a connection.

Consider a vector field $X_c = \sum_i c_i X_i$ having periodic orbits of period one in $\La$. The time-one flow under $X_c$ defines a map $\Phi : M \to M$ and $\Phi (m) = m$ whenever $m \in \La$. For $p \in U$ not lying on $\Lambda$, in general $\pi [\Phi(p)] \not= \pi (p)$; however we can use $\Gamma$ to take $\Phi (p)$ into a point $p' \in \S_{\pi (p)}$, which we denote as $p' = \Gamma (\Phi (p)) := \Psi (p)$. This map $\Psi : \S_m \to \S_m$ is of course the Poincar\'e-Nekhoroshev map, and satisfies $\Psi (m) = m$ for all $m \in \La$. Condition $N$ will be again that the spectrum of the linearization $A$ of $\Psi$ at points $m \in \La$ does not include 1.

Note that now the natural system of coordinates in $U$ to use is $(\phi,s)$, where the $\phi$ are angular coordinates on $\La = \toro^k$ and the $s$ are geodetic coordinates on $\S_{m(\phi)}$.
We can then proceed as in section 3, and obtain again (the equivalent of) lemmas 1,2 and 3 in this framework, with essentially the same proofs as those given in there. The considerations of section 4 are also immediately extended to the non-hamiltonian framework.

\bigskip

Finally, we note that one could in principles consider the case where there is some invariant manifold $\La$ (not necessarily a torus) and the $k$ vector fields $X_i$ are in involution in the sense of Frobenius, i.e. they span a $k$-dimensional Lie algebra ${\cal G}$, $[X_i,X_j]= c_{ij}^k X_k$, but $\G$ is not abelian, $c_{ij}^k \not\equiv 0$. In this case we are not guaranteed that all cycles in the homology of $\La$ can be realized as vector fields in the algebra ${\cal G}$; other parts of our construction also seem to depend crucially on the commutativity of the flows under the $X_i$, and thus appear not to extend to such a more general case.


\vfill

\begin{thebibliography}{19}


\bibitem{Nek1} N.N. Nekhoroshev, ``The Poincar\'e-Lyapounov-Liouville-Arnol'd theorem'', 
{\it Funct. Anal. Appl.} {\bf 28} (1994), 128-129

\bibitem{Arn1} V.I. Arnold, {\it Mathematical Methods of Classical Mechanics}, Springer, Berlin 1983

\bibitem{BG} D. Bambusi and G. Gaeta, ``On persistence of invariant tori and a theorem by Nekhoroshev'', Preprint 2001

\bibitem{BV} D. Bambusi and D. Vella, ``Quasi periodic breathers in hamiltonian lattices with symmetry'', Preprint 2001

\bibitem{Dui} J.J. Duistermaat, ``On global action-angle coordinates'', {\it Comm. Pure Appl. Math.} {\bf 33} (1980), 687-706

\bibitem{Cor} B. Cordani, {\it The Kepler problem}, forthcoming book, Birkhauser

\bibitem{Kuk} S.B. Kuksin, ``An infinitesimal Liouville-Arnold theorem as a criterion of reducibility for variational hamiltonian equations'', {\it Chaos, Sol. Fract.} {\bf 2} (1992), 259-269

\end{thebibliography}
\end{document}